\documentclass[conference]{IEEEtran}
\IEEEoverridecommandlockouts

\usepackage{cite}
\usepackage{amsmath,amssymb,amsfonts}
\usepackage{algorithmic}
\usepackage{graphicx}
\usepackage{textcomp}
\usepackage{xcolor}
\usepackage{url}
\usepackage{tabularx}
\usepackage{comment}
\usepackage{multirow}
\usepackage{soul}
 \usepackage{booktabs}
\def\BibTeX{{\rm B\kern-.05em{\sc i\kern-.025em b}\kern-.08em
    T\kern-.1667em\lower.7ex\hbox{E}\kern-.125emX}}
\begin{document}

\title{PC-Mix: Partial-Component Audio Spoofing Detection under Mixed Speech and Environmental Sound Conditions\\
}

\author{
\IEEEauthorblockN{
Zhenshan Zhang\IEEEauthorrefmark{1},
Xueping Zhang\IEEEauthorrefmark{1},
Linxi Li\IEEEauthorrefmark{2},
Yechen Wang\IEEEauthorrefmark{2},
Ming Li\IEEEauthorrefmark{3}\textsuperscript{*}
}

\IEEEauthorblockA{
\IEEEauthorrefmark{1}Digital Innovation Research Center, Duke Kunshan University, Kunshan, China
}

\IEEEauthorblockA{
\IEEEauthorrefmark{2}OfSpectrum, Inc., Los Angeles, USA
}

\IEEEauthorblockA{
\IEEEauthorrefmark{3}School of Artificial Intelligence, The Chinese University of Hong Kong, Shenzhen, China
}

\IEEEauthorblockA{
\textsuperscript{*}Corresponding author
}
}
\maketitle

\begin{abstract}
Recent studies on partial audio spoofing mainly focus on studio-recorded speech with temporal localization of spoofed segments. However, these studies often overlook realistic conditions where spoofed and bonafide segments simultaneously coexist across speech and environmental sound components. In this paper, we present PC-Mix, the first dataset for partial-component spoofing detection, where either or both audio components may be partially spoofed. In PC-Mix, bonafide and partially spoofed environmental-sound components are first constructed and mixed with speech signals from an existing partial-spoof dataset, producing audio in which either or both components may be locally manipulated. This design addresses two major gaps in existing partial spoofing benchmarks: the lack of realistic environmental sounds in speech partial spoofing scenarios and the absence of partial spoofing detection for environmental sound components. We further establish standardized evaluation protocols and design a joint learning framework to optimize spoofing detection across speech, environmental sound, and mixed audio. Experiments highlight the increased difficulty introduced by mixed conditions. The results demonstrate that training under matched target conditions is more effective than directly transferring models trained on speech or environmental sound components. 
\end{abstract}

\begin{IEEEkeywords}
Audio Anti-Spoofing, Partial Spoofing Detection, Component-Level Audio Anti-Spoofing, Joint Learning
\end{IEEEkeywords}

\section{Introduction}

Audio anti-spoofing has become an important task in trustworthy speech and audio processing, as increasingly realistic synthetic and manipulated audio continues to pose serious security risks to real-world applications~\cite{wu2015spoofing,wang2020asvspoof2019,liu2023asvspoof2021}.

Spoofed audio can be generated or manipulated using increasingly powerful techniques, including voice conversion methods~\cite{wang25ba_interspeech,ren25b_interspeech}, zero-shot speech synthesis methods~\cite{yang2025palle,li25t_interspeech}, and audio editing methods~\cite{huang25c_interspeech,ungersbock2025sao,lan2026smartdj}, posing significant threats to authentication systems and downstream applications.

Most research~\cite{zhang2026multiapi, zhang2026impact, buera2026combining, truong2026addressing} only focuses on utterance-level spoof, ignoring localized manipulation within an utterance and treating the entire signal as either fully real or fully spoofed.
More recently, attention has shifted toward partial spoofing, where only localized temporal regions of an utterance are manipulated while the rest remains bonafide~\cite{zhang2023partialspoof,yi2022add}.
This scenario is more realistic, as modern audio editing tools enable fine-grained manipulation of selected segments without altering the entire recording.
Existing benchmarks~\cite{zhang2023partialspoof,zhang25g_interspeech,luong2025llamapartialspoof,baser25b_interspeech, cai2024avdeepfake1m,cai2025avdeepfake1mpp} for partial-spoofing and temporal deepfake localization have shown that localized attacks are particularly challenging to detect.

However, these studies predominantly focus on the speech component alone and typically assume clean or studio-recorded conditions, overlooking the role of environmental background sounds~\cite{zhang2026multiapi,zhang2026impact,xie2026atadd,delgado2026presentation}. In real-world scenarios, audio recordings are inherently composed of multiple acoustic components, including foreground speech and environmental sounds.
Recent work has begun to explore component-level spoofing, where spoofing may target either the speech component, the environmental component, or both~\cite{zhang2026compspoof,zhang2026overview_esdd2}.
This formulation reflects practical scenarios where adversaries may manipulate only selected audio components such as speech or environmental sound or possibly both.
Such mixed conditions introduce additional complexity, as detection systems must reason about multiple interacting sources rather than a single homogeneous signal.

Nevertheless, existing component spoofing studies mainly consider component-level manipulation at the utterance level and do not explicitly address temporally localized spoofing within each component. To address this gap, we establish a partial component spoof evaluation protocol. We propose PC-Mix, a new dataset for partial-component spoofing under mixed speech and environmental sound conditions. In PC-Mix, we construct bonafide and partially spoofed environmental-sound components and mix them with speech signals from an existing partial-spoof dataset. Each resulting sample may contain localized spoofing in the environmental sound component, the speech component, or both. Original recordings are also included to distinguish naturally recorded audio from constructed mixtures. A joint learning framework is further proposed for frame-level partial-component spoofing detection. The framework adopts a two-stage training strategy, first learning component-specific cues for speech spoofing, environmental-sound spoofing, and original-mix discrimination, and then jointly optimizing them under a unified partial-component spoofing objective. This design combines component-level localization with mix-level scene discrimination.

Our main contributions are summarized as follows:

\begin{itemize}
\item We introduce partial-component spoofing detection as a new evaluation task for partially edited audio, where localized spoofing may occur in speech, environmental sound, or both within a mixed acoustic scene.
\item We propose PC-Mix, the first dataset for partial-component spoofing detection and the first benchmark dataset that includes partially spoofed environmental sounds. PC-Mix addresses two major gaps in existing partial spoofing benchmarks: the lack of realistic environmental sounds in speech partial spoofing scenarios and the absence of partial spoofing detection for environmental sound components.

\item We propose a jointly trained baseline framework for partial-component spoofing detection. The framework performs speech-component spoofing detection, environmental-sound spoofing detection, and original-mix discrimination within a unified model.
\item The source code and dataset are publicly released\footnote{\url{https://anonymous.4open.science/r/PC-Mix-3AFE/}} to facilitate reproducible research and future studies.

\end{itemize}

\section{PC-Mix Dataset}

PC-Mix is designed to study partial-component spoofing under mixed speech and environmental sound conditions. It contains two types of data: original samples and constructed mixed samples. The original samples are real-world recordings that naturally contain speech and environmental sounds. The constructed mixed samples are generated by mixing partially spoofed speech utterances with partially spoofed environmental sound components. 

\subsection{Audio Source}
\subsubsection{Original recorded audio}
Original recordings are included as a reference domain for naturally recorded speech--environmental sound scenes. These samples are not created through artificial mixing, representing real acoustic conditions where speech and environmental sounds coexist. In PC-Mix, original recordings are collected from VGGSound~\cite{chen2020vggsound} and SBCASE~\cite{dubois2005sbcsae}. They are used to help distinguish naturally recorded audio from constructed speech--environmental sound mixtures.

\subsubsection{Partially spoofed speech}
For the speech component, audio is from PartialSpoof~\cite{zhang2023partialspoof}. 

\subsubsection{Partially spoofed environmental sound}
\begin{table}[!t]
\centering
\caption{Statistics of the environmental partial-spoof components.}
\label{tab:env_partial_stats}
\scriptsize
\setlength{\tabcolsep}{1.6pt}
\renewcommand{\arraystretch}{1.08}
\begin{tabularx}{\columnwidth}{@{}lrrrllXl@{}}
\toprule
\textbf{Split} & \textbf{\#All} & \textbf{\#Bona} & \textbf{\#Spoof} 
& \textbf{BG} & \textbf{Event Source} & \textbf{Fusion} & \textbf{Shift} \\
\midrule
Train & 25,380 & 12,715 & 12,665
& SONYC & ALDM2/US8K & Duck. & Train \\
\midrule
E0 & 17,812 & 8,931 & 8,881
& SONYC & ALDM2/FSD50K & Duck. & ID \\

E1 & 14,248 & 7,181 & 7,067
& SONYC & AudioGen & Duck. & Gen. OOD \\

E2 & 14,248 & 7,045 & 7,203
& SONYC & ALDM2 & Cross. & Fus. OOD \\

E3 & 17,809 & 8,757 & 9,052
& DEMAND & ALDM2 & Duck. & BG OOD \\

E4 & 7,125 & 3,562 & 3,563
& WHAM! & ALDM2 & Duck. & Noise OOD \\
\bottomrule
\end{tabularx}

\vspace{0.5mm}
\begin{minipage}{0.96\columnwidth}
\scriptsize
\textit{Note:} BG denotes the background source. OOD denotes the out-of-domain setting. ID denotes the in-domain setting. Duck., Cross., ALDM2, and US8K denote ducking overlay, crossfade-based fusion, AudioLDM2, and UrbanSound8K, respectively.
\end{minipage}
\end{table}

This work introduces the first environmental-sound partial spoofing dataset designed for component-level spoofing analysis. For the environmental sound component, partially spoofed samples are constructed by inserting real or generated event sounds into real environmental background recordings.

In the training set, SONYC~\cite{cartwright2020sonycustv2} is used as the background source. Spoofed event sounds are generated by AudioLDM2~\cite{liu2024audioldm2} and real events are curated from UrbanSound8K~\cite{salamon2014urbansound}. The event sounds are inserted into background recordings using a controlled ducking-overlay strategy~\cite{torcoli2019preferred}, where the background is locally attenuated during the inserted event. This preserves acoustic overlap while keeping the event perceptually salient.

\begin{table}[t]
\centering
\caption{PC-Mix class definitions for utterance-level and frame-level evaluation.}
\label{tab:pcmix_class_definitions}
\scriptsize
\setlength{\tabcolsep}{2.2pt}
\renewcommand{\arraystretch}{1.05}
\begin{tabularx}{\columnwidth}{@{}c c c c c X@{}}
\toprule
\textbf{Level} & \textbf{ID} & \textbf{Mixed} & \textbf{Speech} & \textbf{Env.} & \textbf{Description} \\
\midrule
\multirow{5}{*}{Utt.}
& 0 & No  & --        & --        & Original outdoor recording \\
& 1 & Yes & bonafide & bonafide & bonafide speech with bonafide env. \\
& 2 & Yes & Spoofed   & bonafide & spoofed speech with bonafide env. \\
& 3 & Yes & bonafide & Spoofed   & bonafide speech with spoofed env. \\
& 4 & Yes & Spoofed   & Spoofed   & spoofed speech with spoofed env. \\
\midrule
\multirow{4}{*}{Frame}
& 0 & Yes & bonafide & bonafide & bonafide speech and env. frame \\
& 1 & Yes & Spoofed   & bonafide & spoofed speech frame \\
& 2 & Yes & bonafide & Spoofed   & spoofed env. frame \\
& 3 & Yes & Spoofed   & Spoofed   & spoofed speech and env. frame \\
\bottomrule
\end{tabularx}
\end{table}
In the evaluation set, five evaluation subsets are constructed, denoted as E0--E4. E0 serves as the in-domain evaluation set. It uses SONYC backgrounds, together with a new event pool composed of non-overlapping events generated by AudioLDM2 and real events from FSD50K~\cite{fonseca2022fsd50k}. E1 evaluates generator shift by replacing AudioLDM~2 with AudioGen~\cite{kreuk2023audiogen}. E2 evaluates fusion shift by replacing the default ducking-overlay strategy with hard replacement followed by crossfade~\cite{negroni2024splicing}, while keeping the source domains unchanged. E3 evaluates background-domain shift by replacing SONYC with DEMAND~\cite{thiemann2013demand}. E4 evaluates noise-domain shift by using WHAM! noise as the background source~\cite{wichern2019wham}, resulting in more noise-dominated acoustic conditions. The statistics of the training and evaluation splits are summarized in Table~\ref{tab:env_partial_stats}.

\begin{table}[t]
\centering
\caption{Utterance-level class distribution of PC-Mix.}
\label{tab:pcmix_utt_distribution}
\scriptsize
\setlength{\tabcolsep}{3.2pt}
\renewcommand{\arraystretch}{1.08}
\begin{tabular}{lrrrrrrr}
\toprule
\textbf{Split} & \textbf{Orig} & \textbf{BB} & \textbf{BS} & \textbf{SB} & \textbf{SS} & \textbf{\#Total} & \textbf{Duration.} \\
\midrule
Train & 12,155 & 1,287 & 1,293 & 11,427 & 11,373 & 37,535 & 98.95h \\
Eval  & 17,809 & 3,703 & 3,666 & 31,773 & 32,100 & 89,051 & 41.71h \\
\midrule
Total & 29,964 & 4,990 & 4,959 & 43,200 & 43,473 & 126,586 & 140.65 h \\
\bottomrule
\end{tabular}

\vspace{0.5mm}
\begin{minipage}{0.96\columnwidth}
\scriptsize
\textit{Note:} Orig denotes original recordings. 
For mixed samples, the first and second letters denote the speech and environmental-sound labels, respectively. 
B and S denote bonafide and spoofed.
\end{minipage}
\end{table}
\subsection{Mix Process}

Each partially spoofed speech utterance is paired with a partially spoofed environmental-sound recording and inserted at a random position on its timeline, with the output duration kept equal to the environmental-sound component by zero-padding shorter speech or truncating longer speech. All audio is sampled at 16 kHz. Training and evaluation mixtures are generated with nominal speech-to-environmental-sound SNRs of 8 dB and 10 dB, respectively, using fixed speech and environmental-sound gains of 1.0 and 1.4 with clipping protection; the effective SNR may vary slightly because environmental-sound scaling is computed for each pair.

PC-Mix provides utterance-level and frame-level labels for both speech and environmental-sound components. Speech labels are inherited from PartialSpoof, while environmental-sound labels are derived from the constructed environmental-sound component. Segment labels are provided at 40, 80, 160, 320, and 640 ms resolutions. In addition, original-mix labels are also provided to distinguish naturally recorded audio from constructed mixtures. Based on these annotations, PC-Mix defines five utterance-level classes and four frame-level classes, as summarized in Table~\ref{tab:pcmix_class_definitions}; frame-level classes are defined only for constructed mixtures.
\begin{figure}[!t]
    \centering
    \includegraphics[width=1\linewidth]{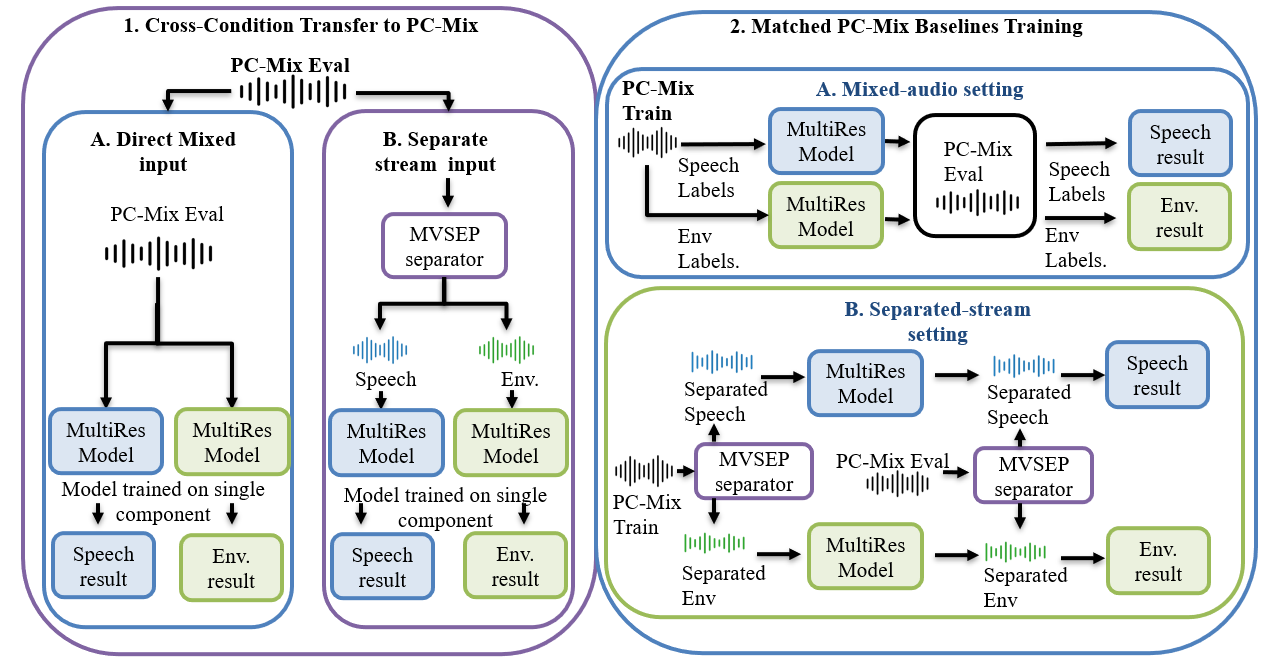}
    \caption{Experimental pipeline for the baseline studies. Single-component detectors and matched PC-Mix trained models are evaluated on the PC-Mix evaluation set. Original recordings are excluded from these component-level baselines.}
    \label{fig:exp_pipeline}
\end{figure}

Overall, PC-Mix contains 126,586 clips and 140.65 hours of audio. The utterance-level class distribution of the training and evaluation splits is reported in Table~\ref{tab:pcmix_utt_distribution}.

\section{Methodology}
\subsection{Baseline System}

We conduct two types of PC-Mix baseline evaluations, as illustrated in Fig.~\ref{fig:exp_pipeline}. Original real-world recordings in PC-Mix are excluded from these component-level baseline experiments because they do not provide separate speech- or environmental-sound spoofing annotations; they are later reintroduced in the joint learning experiments. 

\subsubsection{Cross-Condition Transfer to PC-Mix}
We trained single-component speech and environmental-sound detectors on PartialSpoof and the constructed environmental-sound data, respectively. Then we transferred the single-component detectors to PC-Mix evaluation. During evaluation, two input settings are considered. In the mixed-audio input setting, both detectors are directly evaluated on the PC-Mix mixtures using their corresponding component labels. In the separated-stream input setting, each PC-Mix waveform is first separated into speech and environmental-sound streams using the MVSEP separator~\cite{uhlich2024sound}, after which each detector is evaluated on its corresponding separated stream.

\subsubsection{Matched PC-Mix Baselines}
We train matched PC-Mix detectors under two input settings. In the mixed-audio input setting, speech-focused and environmental-sound-focused detectors are trained on PC-Mix training mixtures and evaluated on PC-Mix evaluation mixtures using the corresponding component labels. In the separated-stream input setting, both the training and evaluation mixtures are first separated into speech and environmental-sound streams using MVSEP~\cite{uhlich2024sound}; each detector is then trained and evaluated on its corresponding separated stream. These matched baselines provide references for assessing the effect of matched mixed-condition training and for comparing with the proposed joint learning framework.

\subsection{Three-head Joint Learning Framework}

\begin{figure*}[t]
    \centering
    \includegraphics[width=\textwidth]{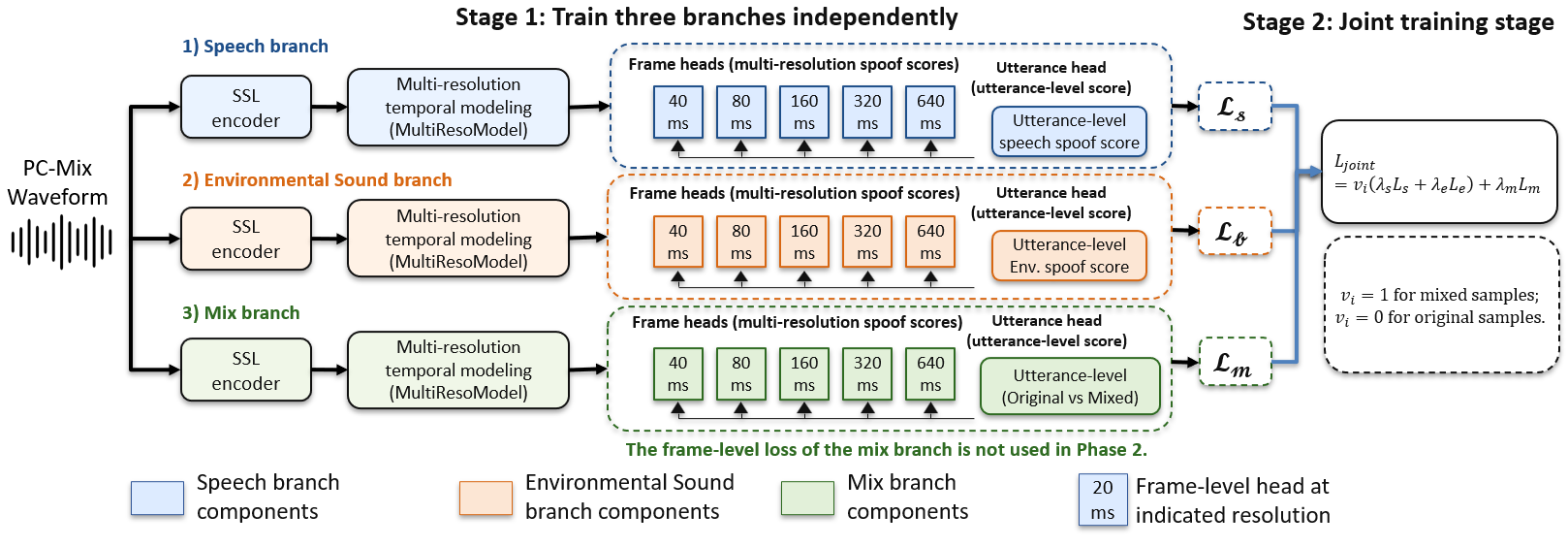}
    \caption{Two-stage three-head joint learning framework for partial-component spoofing detection. Stage 1 independently trains speech, environmental-sound, and mix branches with their corresponding labels on PC-Mix train. Stage 2 jointly optimizes the pretrained branches, applying component-level losses only to constructed mixtures and utterance-level original-mix supervision to the mix branch. Env denotes Environmental sounds.}
    \label{fig:three_head_framework}
\end{figure*}

\subsubsection{Framework Architecture}
A three-head joint learning framework is established for partial-component spoofing detection, as illustrated in Fig.~\ref{fig:three_head_framework}. The framework consists of three MultiResoModel~\cite{luong2025llamapartialspoof} branches: a speech branch, an environmental sound branch, and a mix branch. The speech and environmental sound branches are tailored to detect partial spoofing in the corresponding audio components, while the mix branch is designed to distinguish constructed speech--environmental sound mixtures from original real-world audio recordings. Each branch follows the same multi-resolution architecture, where SSL-based acoustic representations are processed by temporal modeling layers and multiple detection heads. During joint training, the three branches are optimized in a unified loss, allowing component-level spoofing cues and mix-level detection cues to be learned simultaneously.

Given an input waveform $\mathbf{x}$, the three branches produce:
\begin{equation}
    \mathbf{z}^{s} = f_s(\mathbf{x}), \quad
    \mathbf{z}^{e} = f_e(\mathbf{x}), \quad
    \mathbf{z}^{m} = f_m(\mathbf{x}),
\end{equation}
where $f_s(\cdot)$, $f_e(\cdot)$, and $f_m(\cdot)$ denote the speech, environmental sound, and mix branches, respectively. Their outputs $\mathbf{z}^{s}$, $\mathbf{z}^{e}$, and $\mathbf{z}^{m}$ contain multi-resolution frame-level prediction scores and an utterance-level prediction score. The speech and environmental sound branch predict partial spoofing in the corresponding components, while the mix branch predicts whether the input audio is an original recording or a constructed mix. 

\subsubsection{Two-Stage Training Strategy}
The framework is trained in two stages. In the first stage, the three branches are trained independently using their corresponding labels. The speech branch is trained with speech-component partial spoofing labels, the environmental sound branch is trained with environmental-sound partial spoofing labels, and the mix branch is trained with original-mix labels. In the second stage, the three pretrained branches are combined into a single joint framework and optimized together. For the mix branch, only the utterance-level original-mix loss is used during second-stage joint training, while its frame-level outputs are excluded from loss supervision and not used for reported metrics. This design allows each branch to first learn component-specific spoofing cues and then adapt jointly under the partial-component spoofing setting.

\subsubsection{Joint Training}

Let $\mathcal{K}$ denote the set of frame-level temporal resolutions. 
For $i$-th sample, the speech branch output consists of frame-level scores 
$\{\mathbf{z}^{s}_{i,k}\}_{k\in\mathcal{K}}$ and an utterance-level score 
$z^{s}_{i,u}$. Similarly, the environmental-sound branch produces 
$\{\mathbf{z}^{e}_{i,k}\}_{k\in\mathcal{K}}$ and $z^{e}_{i,u}$, and the mix branch 
produces $\{\mathbf{z}^{m}_{i,k}\}_{k\in\mathcal{K}}$ and $z^{m}_{i,u}$.

The corresponding speech and environmental-sound labels are denoted as 
$\{\mathbf{y}^{s}_{i,k}\}_{k\in\mathcal{K}}$, $y^{s}_{i,u}$, 
$\{\mathbf{y}^{e}_{i,k}\}_{k\in\mathcal{K}}$, and $y^{e}_{i,u}$. 
For the mix branch, $y^{m}_{i,u}$ denotes the utterance-level original-mix label, 
where $y^{m}_{i,u}=1$ indicates a constructed mix and $y^{m}_{i,u}=0$ indicates an original sample. 
During first-stage independent training of the mix branch, 
$\mathbf{y}^{m}_{i,k}$ denotes the corresponding frame-level original-mix label at resolution $k$. 
The P2SGrad loss\cite{wang2021comparative} is used for all binary detection branches.

For the speech branch, the multi-resolution loss is defined as
\begin{equation}
\begin{split}
    \mathcal{L}^{s}_{i}
    =
    \sum_{k \in \mathcal{K}}
    \ell(\mathbf{z}^{s}_{i,k}, \mathbf{y}^{s}_{i,k})
    +
    \ell(z^{s}_{i,u}, y^{s}_{i,u}).
\end{split}
\end{equation}
The environmental sound branch follows the same form:
\begin{equation}
\begin{split}
    \mathcal{L}^{e}_{i}
    =
    \sum_{k \in \mathcal{K}}
    \ell(\mathbf{z}^{e}_{i,k}, \mathbf{y}^{e}_{i,k})
    +
    \ell(z^{e}_{i,u}, y^{e}_{i,u}).
\end{split}
\end{equation}

For the mix branch, first-stage independent training uses both frame-level and utterance-level supervision:
\begin{equation}
\begin{split}
    \mathcal{L}^{m,\mathrm{P1}}_{i}
    =
    \sum_{k \in \mathcal{K}}
    \ell(\mathbf{z}^{m}_{i,k}, \mathbf{y}^{m}_{i,k})
    +
    \ell(z^{m}_{i,u}, y^{m}_{i,u}).
\end{split}
\end{equation}
During second-stage joint training, only the utterance-level original-mix loss is used:
\begin{equation}
    \mathcal{L}^{m,\mathrm{P2}}_{i}
    =
    \ell(z^{m}_{i,u}, y^{m}_{i,u}).
\end{equation}

Since original samples do not contain separated speech and environmental-sound spoofing annotations, component-level losses are applied only to constructed mixtures. This is implemented as
\begin{equation}
    v_i =
    \begin{cases}
    1, & y^{m}_{i,u}=1, \\
    0, & y^{m}_{i,u}=0.
    \end{cases}
\end{equation}

The final joint objective over a mini-batch $\mathcal{B}$ is
\begin{equation}
    \mathcal{L}_{\mathrm{joint}}
    =
    \frac{1}{|\mathcal{B}|}
    \sum_{i \in \mathcal{B}}
    \left(
    \lambda_s v_i \mathcal{L}^{s}_{i}
    +
    \lambda_e v_i \mathcal{L}^{e}_{i}
    +
    \lambda_m \mathcal{L}^{m,\mathrm{P2}}_{i}
    \right).
\end{equation}
where $\lambda_s$, $\lambda_e$, and $\lambda_m$ are the loss weights for the speech, environmental sound, and mix branches. Thus, mix-level supervision is applied to all samples, while speech and environmental sound losses are masked out for original recordings without component-level annotations.

\section{Experiments}

\subsection{Experimental Settings}

\subsubsection{Data Processing}

Audio samples are resampled to 16 kHz and processed as 9.6-second segments with a trailing padding duration of 0.005 s. All models use 160 ms frame-level supervision and produce frame-level outputs at the same resolution, together with an utterance-level output. Random seeking and mask-based training are enabled. 

\subsubsection{Training Configuration}
All detectors are implemented with MultiResoModel~\cite{luong2025llamapartialspoof} as the backbone. All models are optimized using Adam with a learning rate of $1\times10^{-5}$. A StepLR scheduler is applied with a step size of 10 epochs and a decay factor of 0.5. 

Equal branch weights are used in all joint-training experiments, i.e., $\lambda_s=\lambda_e=\lambda_m=1$. 
During joint training, the mix branch is trained on all samples, while the speech and environmental-sound branches are optimized only when valid component-level annotations are available. 

\subsubsection{Evaluation Metrics}

Performance is evaluated using both binary and multi-class metrics. For the baseline evaluations, only equal error rate (EER) is reported at both the utterance and frame levels. For each binary branch in the joint framework, utterance-level metrics include EER, F1 score, and accuracy. For the speech and environmental sound branches, the same binary metrics are also computed at the frame level. Branch predictions are further mapped to the class definitions in Table~\ref{tab:pcmix_class_definitions}. At the utterance level, five-class classification is evaluated with a mix-gated decision rule. Accuracy, macro F1, and weighted F1 are reported for this evaluation. At the frame level, four-class classification is evaluated on mixed utterances based on the authenticity states of speech and environmental sounds, using the same multi-class metrics. Since original-mix supervision is defined at the recording level, the mix-branch frame-level outputs are not used in joint evaluation; therefore, frame-level five-class classification is not reported.

\subsection{Experimental results}

\subsubsection {Single-Component Partial Spoofing Detection Results}
\begin{table}[!t]
\centering
\caption {Single-component partial-spoofing speech/env detector trained and evaluated on speech and environmental-sound eval set.}
\label{tab:single_component}
\scriptsize
\setlength{\tabcolsep}{3.5pt}
\renewcommand{\arraystretch}{1.05}
\begin{tabular}{llccc}
\toprule
\textbf{Train data} & \textbf{Eval data} & \textbf{Epoch} & \textbf{Utt. EER (\%)} & \textbf{Frame EER (\%)} \\
\midrule
Speech train & Speech eval & 25 & 2.16 & 11.36 \\
Env. sound train & Env. sound eval & 25 & 3.55 & 5.89 \\
\bottomrule
\end{tabular}

\vspace{0.4mm}
\vspace{-1.5mm}
\end{table}

\begin{table}[!t]
\centering
\caption{Environmental-sound partial-spoofing detection performance across different in-domain and out-of-domain evaluation subsets.}
\label{tab:env_sound_subsets}
\scriptsize
\setlength{\tabcolsep}{3.5pt}
\renewcommand{\arraystretch}{1.05}
\begin{tabular}{llccc}
\toprule
\textbf{Train data} & \textbf{Eval data} & \textbf{Epoch} & \textbf{Utt. EER (\%)} & \textbf{Frame EER (\%)} \\
\midrule
Env. sound train & Env. sound eval E0 & 25 & 0.67 & 2.47 \\
Env. sound train & Env. sound eval E1 & 25 & 1.88 & 7.98 \\
Env. sound train & Env. sound eval E2 & 25 & 0.53 & 2.57 \\
Env. sound train & Env. sound eval E3 & 25 & 9.15 & 13.04 \\
Env. sound train & Env. sound eval E4 & 25 & 3.69 & 3.93 \\
\bottomrule
\end{tabular}

\vspace{0.4mm}
\vspace{-1.5mm}
\end{table}
We train MultiResoModel-based speech and environmental-sound spoofing detectors on PartialSpoof and the constructed partial-spoof environmental-sound training data, respectively, and evaluate them on the corresponding single-component test sets. Table~\ref{tab:single_component} shows that both speech and environmental-sound spoofing detectors perform well under matched conditions, indicating that each detection head is effective on its corresponding component.

Table~\ref{tab:env_sound_subsets} presents the evaluation results of the environmental-sound detector on the proposed E0--E4 subsets. The results show that these subsets expose different challenges for partial environmental-sound spoofing detection. The detector performs well on the in-domain E0 set and remains stable under fusion shift in E2, suggesting that inserted spoofed events can be localized when the background domain is consistent. However, performance degrades most under background-domain shift in E3, indicating that environmental context mismatch is a key challenge. These results demonstrate that the proposed subsets provide a useful benchmark for evaluating localized spoofing detection under different domain shifts.

\subsubsection{Baseline result}

\begin{table}[t]
\centering
\caption{Cross-condition transfer evaluation from the single-component speech/env trained detector to PC-Mix mixed-audio and separated speech or environment sound inputs.}
\label{tab:cross_condition}
\scriptsize
\setlength{\tabcolsep}{2.2pt}
\renewcommand{\arraystretch}{1.08}
\begin{tabular}{@{}llcccc@{}}
\toprule
\textbf{Setting} & \textbf{Target} & \textbf{Input} & \textbf{Epoch} & \textbf{Utt. EER (\%)} & \textbf{Frame EER (\%)} \\
\midrule
Mixed & Speech & Mixed audio & 25 & 29.35 & 32.88 \\
Mixed & Env. sound & Mixed audio & 25 & 44.80 & 33.44 \\
\midrule
Separated & Speech & Sep. speech & 25 & 51.38 & 37.51 \\
Separated & Env. sound & Sep. env. & 25 & 22.89 & 17.65 \\
\bottomrule
\end{tabular}

\vspace{0.6mm}
\parbox{0.96\linewidth}{\scriptsize
\textit{Note:} Env. sound denotes environmental sound, and Utt. denotes utterance-level.
}
\end{table}

This experiment corresponds to the cross-condition transfer setting, where single-component detectors are evaluated directly under PC-Mix conditions. As shown in Table~\ref{tab:cross_condition}, both input settings suffer substantial degradation compared to the single-component results in Table~\ref{tab:single_component}. This indicates that detectors trained only on speech or environmental sounds cannot be directly transferred to PC-Mix conditions. Although separated-stream evaluation improves environmental-sound detection over mixed-audio evaluation, it remains far from the single-component setting and further degrades speech detection.

\begin{table}[t]
\centering
\caption{Matched PC-Mix baselines for component-specific spoofing detection, where baseline models are trained and evaluated using either the complete PC-Mix train audio or the separated speech and environmental-sound streams.}
\label{tab:matched_mixed}
\scriptsize
\setlength{\tabcolsep}{3.5pt}
\renewcommand{\arraystretch}{1.08}
\begin{tabular}{llccc}
\toprule
\textbf{Train \& eval setting} & \textbf{Target} & \textbf{Epoch} & \textbf{Utt. EER (\%)} & \textbf{Frame EER (\%)} \\
\midrule
Mixed-audio input & Speech & 30 & 8.53 & 22.52 \\
Mixed-audio input & Env. sound & 30 & 3.57 & 5.68 \\
\midrule
Separated-stream input & Speech & 30 & 7.83 & 22.61 \\
Separated-stream input & Env. sound & 30 & 7.43 & 8.21 \\
\bottomrule
\end{tabular}

\vspace{0.6mm}
\parbox{0.96\linewidth}{\scriptsize
\textit{Note:} Env. sound denotes environmental sound, and Utt. denotes utterance-level.
}
\end{table}

Table~\ref{tab:matched_mixed} reports the matched PC-Mix baseline results, where component-specific detectors are trained and evaluated under the same PC-Mix input setting. Compared with cross-condition transfer in Table~\ref{tab:cross_condition}, matched PC-Mix training greatly improves performance. In the mixed-audio setting, speech utterance-level EER decreases from 29.35\% to 8.53\%, and environmental-sound utterance-level EER decreases from 44.80\% to 3.57\%. This confirms that the cross-condition degradation mainly comes from the mismatch between single-component training and mixed acoustic conditions.
\begin{table*}[!t]
\centering
\caption{Ablation comparison between the assembled independently trained branches and the jointly trained three-head model. Both models are trained on the PC-Mix training set and evaluated on the PC-Mix evaluation set. ``Pre-joint'' denotes the assembled model without second-stage joint optimization, and ``Joint-E8'' denotes the model after eight epochs of joint training.}
\label{tab:joint_training_comparison}
\small
\setlength{\tabcolsep}{5pt}
\begin{tabular}{llccc}
\toprule
\multicolumn{5}{l}{\textit{(a) Binary head performance}} \\
\midrule
Level & Head & EER $\downarrow$ & F1 $\uparrow$ & ACC $\uparrow$ \\
\midrule
\multirow{3}{*}{Utterance}
& Speech & 8.72 $\rightarrow$ 7.86 {\scriptsize (-0.86)} & 94.56 $\rightarrow$ 94.58 {\scriptsize (+0.02)} & 89.69 $\rightarrow$ 89.72 {\scriptsize (+0.03)} \\
& Env. & 3.59 $\rightarrow$ 3.12 {\scriptsize (-0.47)} & 76.41 $\rightarrow$ 92.67 {\scriptsize (+16.26)} & 69.36 $\rightarrow$ 92.17 {\scriptsize (+22.81)} \\
& Mix & 0.96 $\rightarrow$ 0.42 {\scriptsize (-0.54)} & 99.43 $\rightarrow$ 99.22 {\scriptsize (-0.21)} & 99.10 $\rightarrow$ 98.77 {\scriptsize (-0.33)} \\
\midrule
\multirow{2}{*}{Frame}
& Speech & 22.54 $\rightarrow$ 21.84 {\scriptsize (-0.70)} & 71.70 $\rightarrow$ 72.55 {\scriptsize (+0.85)} & 77.27 $\rightarrow$ 78.07 {\scriptsize (+0.80)} \\
& Env. & 5.18 $\rightarrow$ 3.87 {\scriptsize (-1.31)} & 89.93 $\rightarrow$ 90.80 {\scriptsize (+0.87)} & 96.94 $\rightarrow$ 96.96 {\scriptsize (+0.02)} \\
\midrule
\multicolumn{5}{l}{\textit{(b) Multi-class decision performance}} \\
\midrule
Task &  & ACC $\uparrow$ & Macro F1 $\uparrow$ & Weighted F1 $\uparrow$ \\
\midrule
Utterance 5-class &  & 69.40 $\rightarrow$ 85.12 {\scriptsize (+15.72)}
& 44.80 $\rightarrow$ 54.31 {\scriptsize (+9.51)}
& 64.78 $\rightarrow$ 81.52 {\scriptsize (+16.74)} \\
Frame 4-class &  & 74.92 $\rightarrow$ 75.72 {\scriptsize (+0.80)}
& 71.58 $\rightarrow$ 72.58 {\scriptsize (+1.00)}
& 75.16 $\rightarrow$ 76.04 {\scriptsize (+0.88)} \\
\bottomrule
\end{tabular}

\vspace{1mm}
\begin{minipage}{0.98\textwidth}
\footnotesize
\textit{Note:} In each entry, the value before $\rightarrow$ is the Pre-joint result and the value after ``$\rightarrow$'' is the Joint-E8 result. Values in parentheses indicate absolute changes after joint training. The mix branch is evaluated only at the utterance level for original-mix discrimination. Env. denotes Environmental sound. All values are in \%.
\end{minipage}
\end{table*}
The separated-stream setting also improves over cross-condition transfer, especially for environmental-sound detection, where utterance-level EER decreases from 22.89\% to 7.43\%. However, it does not consistently outperform mixed-audio training: speech results are close across the two settings, while environmental-sound detection is clearly worse after separation. This suggests that source separation may introduce artifacts or remove environmental-sound spoofing cues. Therefore, the proposed joint learning framework adopts the mixed-audio input setting rather than relying on explicit source separation.

\subsubsection{Ablation Study}

Table~\ref{tab:joint_training_comparison} presents an ablation study of the second-stage joint optimization by comparing the jointly trained model with an assembled model composed of the three independently trained branches from stage 1. In this baseline, the independently trained speech and environmental-sound branches from the previous mixed-audio input experiment, together with the independently trained mix branch, are directly assembled and evaluated without second-stage joint optimization. Each entry is reported as ``Pre-joint $\rightarrow$ Joint-E8'', where Pre-joint denotes the assembled stage-one baseline and Joint-E8 denotes the model after eight epochs of joint training.

The results show that joint training is more effective than directly assembling independently trained heads. The improvement is most pronounced for the environmental-sound branch, whose utterance-level F1 and accuracy increase from 76.41\% and 69.36\% to 92.67\% and 92.17\%, respectively. This suggests that environmental-sound spoofing detection is particularly sensitive to the final mixed-audio training condition, where spoofing cues are entangled with speech and other acoustic components. In contrast, the speech branch shows smaller but consistent gains, indicating that it is already relatively stable after stage-one training and mainly benefits from further refinement during joint optimization. The mix branch also maintains strong original-mix discrimination, with a slight decrease in EER, although its F1 and accuracy slightly decrease, suggesting that its operating point may require further calibration.

The multi-class results further confirm the benefit of joint training. At the utterance level, 5-class accuracy increases from 69.40\% to 85.12\%, and weighted F1 also improves substantially, indicating better consistency among the mix, speech, and environmental-sound branches. Macro F1 also increases, although it remains lower than weighted F1 due to the imbalanced class distribution, especially for the smaller mixed classes. This suggests that joint training improves overall decision consistency while partially mitigating the effect of class imbalance. At the frame level, the gains are smaller but still positive, with accuracy, macro F1, and weighted F1 all increasing after joint training. This indicates that frame-level localization remains more difficult than utterance-level classification, since the model must identify short spoofed regions under overlapping speech and environmental-sound components. Overall, joint training mainly strengthens utterance-level component reasoning, while fine-grained temporal localization remains challenging under mixed-audio conditions.

\section{Conclusion}

In this paper, we introduce partial-component spoofing detection as a new evaluation task for partially edited audio. In this setting, spoofing may occur in speech, environmental sounds, or both within a mixed acoustic scene.
We propose PC-Mix, a benchmark dataset for this task. It is, to the best of our knowledge, the first dataset that includes partially spoofed environmental sounds. It also addresses two key gaps in prior benchmarks: the lack of realistic environmental sounds in speech-based partial spoofing and the absence of evaluation for environmental sound spoofing.
We further propose a jointly trained baseline framework for this task. It performs speech-component spoofing detection, environmental-sound spoofing detection, and original-mix discrimination within a unified model.

\section*{Acknowledgment}
The authors used OpenAI ChatGPT only for language editing, grammar checking, and clarity improvement across the manuscript. The tool was not used to generate research ideas, experimental designs, datasets, experimental results, figures, tables, or conclusions. The authors reviewed and verified all content, claims, and results, and remain fully responsible for the manuscript.
\bibliographystyle{IEEEtran}
\bibliography{reference}

\end{document}